\documentclass[preprint,showpacs,preprintnumbers,amsmath,amssymb]{revtex4}

\usepackage{graphicx}% Include figure files
\usepackage{dcolumn}% Align table columns on decimal point
\usepackage{bm}% bold math
\usepackage{slashed}
\begin{document}

%\date{\today}

\title{Classical strings in $AdS_4\times\mathbb{CP}^3$ with three angular momenta}

\author{Sergio Giardino}
 \email{jardino@fma.if.usp.br}
\affiliation{Departamento de F\'{i}sica Matem\'{a}tica. Instituto de F\'{i}sica. Universidade de S\~{a}o Paulo,\\ CP 66318, 05315-970 S\~{a}o Paulo, SP, Brazil.}
\author{Hector L. Carri\'{o}n}
\email{hectors@ect.ufrn.br}
\affiliation{Escola de Ci\^{e}ncias e Tecnologia, Universidade Federal
  do Rio Grande do Norte,\\ Campus Universitário Lagoa Nova, 59078-970,  Natal, RN, Brazil.}

\begin{abstract}
\noindent In this paper, rotating strings in three directions of 
$AdS_4 \times \mathbb{CP}^3$ geometry are studied; its divergent
energy limit, and conserved charges are also determined. An interpretation
of these configurations as either  giant magnons or spiky strings is discussed. 
\end{abstract}

\maketitle
                %\pacs{11.10.Kk, 11.10.Lm, 11.15.Kc, 11.30.Na, 12.38.Aw}

                %\keywords{confinement, affine Toda coupled to matter, solitons,

                %quantum chromodynamics.}

                %PACS numbers: 04.20.Cv, 03.65.Ta, 04.80.Cc

\section{Introduction}

\setcounter{footnote}{0}

Semi-classical quantization \cite{Gubser:2002tv} is one possible way to test the AdS/CFT
correspondence, as there is currently no full quantization of string theory. The energy spectrum of a specific string solution must be identical
to an anomalous dimension spectrum of an also specific dilation
operator of a dual gauge theory.
More recently, a string state dual to an excitation of an
infinite spin chain gauge operator has been proposed \cite{Hofman:2006xt}.
This solution is called giant magnon and, for a string restricted to the
$\mathbb{R}\times S^2$ subspace of $AdS_5\times S^5$, its dispersion
relation has the characteristic expression of a magnon in a spin chain
\begin{equation} 
E-J =2T \left|\sin\frac{p}{2}\right|, \label{E-J}
\end{equation}
\noindent where {\it E, J } and {\it T} are respectively the energy, the
angular momentum, and the tension of the string. {\it p} is
identified from the angle between the extremes of the giant magnon
and $E,\,J\to\infty$. A solution related to the giant magnons, the
spiky string \cite{Kruczenski:2006pk,Kruczenski:2004wg}, 
also has divergent energy and obeys the dispersion relation 
\begin{equation}
E-T\Delta\phi=2T\left(\frac{\pi}{2}-\theta\right), \label{SS_zero}
\end{equation}
\noindent where {\it T} is the string tension,  $\Delta\phi\to \infty$ is the deficit
angle and $\theta$ is the coordinate where the string is peaked.

String configurations that obey the prescription of a giant magnon were pursued
in $AdS_5\times S^5$, {\it e.g.} \cite{Bobev:2006fg,Kruczenski:2006pk,Dimov:2007ey}, but also in other backgrounds
where the gauge/string correspondence was stated,
for example, Lunin and Maldacena \cite{Bobev:2007bm, Bobev:2006fg,
  Chu:2006ae, Ryang:2005pg, Bobev:2005cz},  Maldacena and
N\'{u}nez \cite{Bobev:2005ng}, and $AdS_4\times\mathbb{CP}^3$ \cite{Hollowood:2009sc,Gaiotto:2008cg,Berenstein:2008dc,Lee:2008ui,Abbott:2009um,Ahn:2010eg,Suzuki:2009sc,Kalousios:2010ne,Biswas:2011wu,Ryang:2008rc,Grignani:2008is,Abbott:2008qd,Hollowood:2009tw,Kalousios:2009mp}.
In $ AdS_4 \times \mathbb{CP}^3$ \cite{Chen:2008qq}, giant magnons
with one \cite{Grignani:2008is} and two \cite{Ryang:2008rc} angular
momenta were found. The solution with three angular momenta has not
yet been found, although it has been predicted \cite{Abbott:2008qd}.

With this work we propose candidates for giant magnons and
spiky strings  with three angular momenta in $ AdS_4 \times
\mathbb{CP}^3$. We call them candidates because the dispersion
relations of these configurations are similar to those found
by \cite{Ryang:2008rc,Kluson:2007qu}, although they do not 
match (\ref{E-J}) and (\ref{SS_zero}) precisely. However,
the prescriptions can be reached in a one dimensional limit of our solutions.

This text is organized as follows: in section (\ref{s1}) we describe 
the motion of strings $ \mathbb{R} \times \mathbb{CP}^3$ using  a
specific ansatz, and we calculate
their conserved charges; in section (\ref{s2}) the divergent energy
solutions and their possible giant magnon and spiky string forms are
constructed. The reasons why these structures do not exactly have the
expected giant magnon dispersion relation are also discussed
in this section. Section (\ref{s3}) is a brief conclusion of our findings.

\section{The classical string \label{s1}} 

We start with the complete $AdS_4\times \mathbb{CP}^3$ metric
\begin{eqnarray}
    ds^2 & = &\frac{R^2}{4}\left(-\cosh^2\rho dt^2 + d\rho^2+\sinh^2\rho d\Omega^2_2\right)+\nonumber \\  
         & + &
         R^2\left[d\xi^2+\cos^2\xi\sin^2\xi\left(d\psi+\frac{1}{2}\cos\theta_1d\phi_1-\frac{1}{2}\cos\theta_2d\phi_2 \right)^2 \right. +\nonumber \\ 
         & + &
         \left. \frac{1}{4}\cos^2\xi\left(d\theta^2_1+\sin\theta^2_1d\phi^2_1\right)+\frac{1}{4}\sin^2\xi\left(d\theta^2_2+\sin\theta^2_2d\phi^2_2\right) \right],
  \label{mtr:inic}
  \end{eqnarray} 
%%%%%%%%%%%%%%%%%
%
% mtr:inic
%
%
\noindent where $R^2=4\pi\sqrt{2\lambda}$ and the ranges of the variables are $\xi\in [0,\pi/2]$, 
$\psi\in[-2\pi,2\pi],$  $\theta_{i=1,2}\in[0,\pi]$ and
$\phi_{i=1,2}\in[0,2\pi]$. The first term of (\ref{mtr:inic}) corresponds to the
$AdS_4$ space and the other to the $\mathbb{CP}^3$ space. 

We are seeking string solutions in the subspace where $\rho=0$, namely  $\mathbb{R}\times\mathbb{CP}^3$. Additionally, the $\theta_{i=1,2}$ will be held constant. In this case, the change of variables $\psi\to(\psi+\frac{1}{2}\cos\theta_1\phi_1-\frac{1}{2}\cos\theta_2\phi_2)$, and $\phi_i\to\sin\theta_i\phi_i$ is equivalent to make
$\theta_i=\pi/2$ in (\ref{mtr:inic}).  So, we are going to consider the {\it ansatz} 
%
%%%%%%%%%%%%%%%%%%%%%%%%%%%%%%%%%%%%%%%%%%%%%%%%%%%%%%%%%%%%%%%%%%%
%                                                                 %
%                                                                 %
%         The calculations are in  ads4cp3_9simp.mw               %
%         and ads4cp3_10simp.mw  maple files                      %
%                                                                 %
%                                                                 %
%%%%%%%%%%%%%%%%%%%%%%%%%%%%%%%%%%%%%%%%%%%%%%%%%%%%%%%%%%%%%%%%%%%
%
%
\begin{equation}
t=\kappa\tau, \qquad \theta_{i=1,2}=\frac{\pi}{2},\qquad\xi=\xi(y), \qquad \psi=\omega\tau+P(y),
\qquad \phi_{i=1,2}=\omega_i \tau + f_i(y)
\label{ansatz:1}
\end{equation}
%%%%%%%%%%%%%%%%%
%
% ansatz:1
%
\noindent where $y=\alpha\sigma + \beta \tau$, and $\alpha$, $\beta$,
$\kappa$, $\omega$, and $\omega_i$ are constants. The Polyakov action  that will be used to (\ref{ansatz:1}) is 
\begin{align}
S=T\int d\sigma d\tau
\frac{1}{4}&\left(-\partial_{a}t\partial^{a} t +  4\partial_{a}\xi\partial^{a}\xi+4\cos^2\xi\sin^2\xi\partial_{a}\psi\partial^{a}\psi+\right. \nonumber\\ 
 &\left. +\cos^2\xi\partial_{a}\phi_1\partial^{a}\phi_1+\sin^2\xi\partial_{a}\phi_2\partial^{a}\phi_2\right),
\label{action:1}
\end{align}
%%%%%%%%%%%%%%%
%
%  action:1
%
\noindent where $T=\sqrt{2\lambda}$ is the string tension, $a=\{0,1\}$, $\partial_0=\partial_{\tau}$, and
$\partial_1=\partial_{\sigma}$. The equations of motion are 
\begin{eqnarray}
&&\partial_a\partial^{a}t=0\\
&&8\partial_{a}\partial^{a}\xi-4\left(\sin^2\xi\cos^2\xi\right)_{\xi}\partial_{a}\psi\partial^{a}\psi-\left(\cos²\xi\right)_{\xi}\partial_{a}\phi_1\partial^{a}\phi_1-\left(\sin²\xi\right)_{\xi}\partial_{a}\phi_2\partial^{a}\phi_2=0
\label{xi:1} \\
&&\partial_{a}\left(\sin^2\xi\cos^2\xi\partial^{a}\psi\right)=0 \label{psi:1} \\
&&\partial_{a}\left(\cos^2\xi\partial^{a}\phi_1\right)=0 \label{phi1:1}\\
&&\partial_{a}\left(\sin^2\xi\partial^{a}\phi_2\right)=0 \label{phi2:1}.
\end{eqnarray}
%%%%%%%%%%%%%%%%%%%%%%%%%%%%%%%
%
% xi:1, psi:1, phi1:1, phi2:1
% 
\noindent
The $\xi$ subscript in (\ref{xi:1}) indicates a derivative with respect to the
variable. The conserved charges of the action (\ref{action:1}) are
\begin{eqnarray}
E&=&\frac{T}{2}\int\limits_{0}^{2\pi}d\sigma\,\dot{t}\label{E:1}\\
J_{\psi}&=&2T\int\limits_{0}^{2\pi}d\sigma\cos^2\xi\,\sin^2\xi\,\dot{\psi}\label{Jpsi:1}\\
J_1&=&\frac{T}{2}\int\limits_{0}^{2\pi}d\sigma\cos^2\xi\,\dot{\phi_1}\label{J1:1}\\
J_2&=&\frac{T}{2}\int\limits_{0}^{2\pi}d\sigma\sin^2\xi\,\dot{\phi_2} \label{J2:1}.
\end{eqnarray}
%%%%%%%%%%%%%%%%%%%%%%%%%%%%%%%%%
%
%
%    E:1, Jpsi:1, J1:1, J2:1
%
%
\noindent
Using  (\ref{ansatz:1}) we integrate (\ref{psi:1}),
(\ref{phi1:1}), and (\ref{phi2:1})  with respect to {\it y } to obtain
\begin{eqnarray}
f_{1\,y}&=&\frac{A_1}{\cos^2\xi}+\frac{\beta\omega_1}{\alpha^2-\beta^2}\nonumber\\
f_{2\,y}&=&\frac{A_2}{\sin^2\xi}+\frac{\beta\omega_2}{\alpha^2-\beta^2} \label{z_y}\\
P_{y}&=&\frac{A_{\psi}}{\sin^2\xi\cos^2\xi}+\frac{\omega\beta}{\alpha^2-\beta^2},\nonumber
\end{eqnarray}
%%%%%%%%%%%%%%%%%%%
%
%
% z_y
%
%
\noindent
where $A_1$, $A_2$ and $A_{\psi}$ are integration constants, and
$f_{i\,y}$ and $P_y$ are first derivatives with respect to $y$. The Virasoro constraints
\begin{equation}
g_{\mu\nu}\partial_{\tau} X^{\mu}\partial_{\sigma}X^{\nu}=0\hspace{5mm} \textnormal{and} \hspace{5mm}
g_{\mu\nu}\left(\partial_{\tau}X^{\mu}\partial_{\tau}X^{\nu} +  \partial_{\sigma}X^{\mu}\partial_{\sigma}X^{\nu} \right)=0, \label{Virasoro}
\end{equation}
%%%%%%%%%%%%%%%%
%
%
%  Virasoro
%
%
%\noindent
%plus the ansatz (\ref{ansatz:1}) give
%
%\begin{eqnarray}
%&&4\xi_{y}^2 +
%4\sin^2\xi\cos^2\xi\left(\frac{\omega}{\beta}+P_{y}\right)P_{y}+\cos^2\xi\left(\frac{\omega_1}{\beta}+f_{1y}\right)f_{1y}
%+\sin^2\xi\left(\frac{\omega_2}{\beta}+f_{2y}\right)f_{2y}=0
%\label{v1:1}\nonumber \\ &&  \\ &&
%4\xi_{y}^2 + 4\sin^2\xi\cos^2\xi\left(\frac{\omega^2+2\omega\beta
%    P_{y}}{\alpha^2+\beta^2}+P_{y}^2\right)+\cos^2\xi\left(\frac{\omega_1^2+2\omega_1\beta f_{1y}}{\alpha^2+\beta^2}+f_{1y}^2\right) + \nonumber\\ &&+ \sin^2\xi\left(\frac{\omega_2^2+ 2\omega_2\beta f_{2y}}{\alpha^2+\beta^2}+f_{2y}^2\right)=\frac{\kappa^2}{\alpha^2+\beta^2} \label{v2:1}
%\end{eqnarray}
%%%%%%%%%%%%%%%%%%
%
% v1:1, v2:1
%
%Subtracting (\ref{v1:1}) from (\ref{v2:1}), and using the integrated equations  (\ref{z_y}), we get
%
\noindent and the ansatz (\ref{ansatz:1}) give
\begin{equation}
4\omega A_{\psi} + \omega_1 A_1 + \omega_2 A_2 + \frac{\kappa^2\beta}{\alpha^2-\beta^2}=0
\label{v1-v2}.
\end{equation}
%%%%%%%%%%%%%%%%%%%%
%
%  v1-v2
%
To work out the sum of the Virasoro constraints, we use the variable
$x=\cos 2\xi$,  the choice
\begin{equation}
\omega_1=\omega_2=\nu\qquad\mbox{and} \qquad A_1=A_2=A,
\end{equation}
\noindent and (\ref{v1-v2}) to get 
%
%\begin{eqnarray}
%&&
%-4\nu^2\left(\alpha^2-\beta^2\right)^2\sin^22\xi\,\xi_{y}^2 =
%\nu^2\omega^2\alpha^2\cos^42\xi+\nonumber\\ &&+\nu^2\left[\left(\alpha^2+\beta^2\right)\kappa^2-\alpha^2\left(\nu^2+2\omega^2\right)\right]\cos^22\xi+16\left(\alpha^2-\beta^2\right)^2\left(\nu^2+\omega^2\right)A_{\psi}^2+
%\nonumber \\ &+&
%8\kappa^2\omega\beta\left(\alpha^2-\beta^2\right)A_{\psi}+\nu^4\alpha^2+\kappa^4\beta^2+\nu^2\left[\alpha^2\omega^2-\left(\alpha^2+\beta^2\right)\kappa^2\right] \label{2xi2:0}.
%\end{eqnarray}
%%%%%%%%%%%%%%%%%%%%
%
%
%    2xi2:0
%
%
%\noindent With the change of variables  $\cos2\xi\to x$ in (\ref{2xi2:0}) we get
%
\begin{equation}
\frac{dx}{\sqrt{\left(x^2-x_{-}^2\right)\left(x^2_{+}-x^2\right)}}=\pm\frac{\omega\alpha}{\alpha^2-\beta^2}dy, \label{x:0}
\end{equation}
\noindent where
%%%%%%%%%%%%%%%%%%%%%%%%%%%%
%
%
%      x:0
%
%
\begin{eqnarray}
x_{\pm}^2=&&\frac{1}{2\omega^2\nu\alpha^2}\Bigg\{\nu\left[\alpha^2\left(\nu^2+2\omega^2\right)-\left(\alpha^2+\beta^2\right)\kappa^2\right]\pm
 \label{x_pm1}\nonumber \\ &&\pm
\bigg[-64\alpha^2\omega^2\left(\alpha^2-\beta^2\right)^2\left(\omega^2+\nu^2\right)A_{\psi}^2-32\omega^3\alpha^2\beta\kappa^2\left(\alpha^2-\beta^2\right)A_{\psi}-\nonumber\\
&&-4\omega^2\alpha^2\beta^2\kappa^4+\nu^2\left[\left(\alpha^2+\beta^2\right)\kappa^2-\nu^2\alpha^2\right]^2\bigg]^{\frac{1}{2}}\Bigg\}. 
\end{eqnarray}
%%%%%%%%%%%%%%
%
%
% x_pm1
%
%
\noindent Using (\ref{z_y}) and (\ref{x:0}) we can
rewrite the conserved charges (\ref{E:1}), (\ref{Jpsi:1}),
(\ref{J1:1}), and (\ref{J2:1}). For appropriate integration limits we
have
\begin{eqnarray}
&&E=\frac{T\kappa}{\omega}\frac{\alpha^2-\beta^2}{\alpha^2}\intop^{x_f}_{x_i}\frac{dx}{\sqrt{(x_+^2-x^2)(x^2-x_-^2)}}\label{Ed1}\\ 
&&J_{i=1,2}=\left(\beta A_i +\frac{1}{2}\frac{\omega_i\alpha^2}{\alpha^2-\beta^2}\right)\frac{E}{\kappa}\mp\frac{T}{4}\frac{\omega_i}{\omega}\intop_{x_i}^{x_f}\frac{xdx}{\sqrt{(x_+^2-x^2)(x^2-x_-^2)}}\label{Jid}\\ 
&&J_{\psi}=\left(4A_{\psi}\beta+\frac{\omega\alpha^2}{\alpha^2-\beta^2}\right)\frac{E}{\kappa}-\frac{T}{2}\intop_{x_i}^{x_f}\frac{x^2dx}{\sqrt{(x_+^2-x^2)(x^2-x_-^2)}}\label{Jpsid},
\end{eqnarray}
%%%%%%%%%
%
% Jid
%
\noindent where the upper sign in (\ref{Jid}) corresponds to $i=1$ and
the bottom sign to $i=2$; this notation is used in the whole paper. We also calculate the deficit angles using the
general formula $\delta\varphi=\int d\varphi$ and the results are
\begin{eqnarray}
&&\Delta\phi_{i=1,2}=\left(2A_i+\frac{\beta\omega_i}{\alpha^2-\beta^2}\right)\alpha\frac{E}{\kappa
T}\mp
2A_i\frac{\alpha^2-\beta^2}{\alpha\omega}\intop_{x_i}^{x_f}\frac{x}{1\pm
x}\frac{dx}{\sqrt{(x_+^2-x^2)(x^2-x_-^2)}}\label{delta_phi}\\ 
&&\Delta\psi=\left(4A_\psi+\frac{\beta\omega}{\alpha^2-\beta^2}\right)\alpha\frac{E}{\kappa
T}+
4A_\psi\frac{\alpha^2-\beta^2}{\alpha\omega}\intop_{x_i}^{x_f}\frac{x^2}{1-
x^2}\frac{dx}{\sqrt{(x_+^2-x^2)(x^2-x_-^2)}}.
\label{delta_psi}
\end{eqnarray}
%%%%%%%%%%%%%%%%%%%%%%%%%%
%
% delta_phi e delta_psi
%
String energy (\ref{Ed1})  can be written in terms
of an elliptic integral of first kind. From  (\ref{x:0}) we get
\begin{equation}
\int\limits_{x_1}^{x_2}\frac{dx}{\sqrt{\left(x^2_{+}-x^2\right)\left(x^2-x^2_{-}\right)}}=g
 F\left(\varphi,\,k\right). \label{E1:0}
\end{equation}
%%%%%%%%%%%%
%
% E1:0
%
\noindent  The parameters $g,\,\varphi$ and $k$ depend on  $x_1,
\,x_2,\,x_+,$ and $x_-$. There are some possibilities of integration according to  
the choice of the integration limits. Excluding the non-physical cases of $x_{\pm}<0$ and $x_{\pm}>1$, 
the results for the general formula (\ref{E1:0})  are arranged in Table~\ref{T1}. 
\vspace{2mm}

\begin{table}[h]
\caption{\label{T1} Integration's cases}
\begin{ruledtabular}
\begin{tabular}{cccccc}
Case&$x_{\pm}$ & Range of  {\it x} &  $g^{-1}$ & $\sin^2\varphi$  & $k^2$ \\ 
\hline
{\bf 1.} & $x_{+}^2>1$ and $x_{-}^2<0$ & $[0,\,1]$  &  $\sqrt{x_{+}^2+\left| x_{-}^2\right|}$ &
$\frac{x_{+}^2+\left| x_{-}^2\right|}{x_{+}^2\left(1+\left| x_{-}^2\right|\right)}$  & 
$\frac{x_{+}^2}{x_{+}^2+|x_{-}^2|}$  \\  
{\bf 2.} & $x_{+}^2>1$ and $x_{-}^2\in [0,\,1]$ & $[x_{-},\,1]$ & $x_{+}$  & $
\frac{x_{+}^2\left(1-x_{-}^2\right)}{x_{+}^2-x_{-}^2} $ &$1-\frac{x_{-}^2}{x_{+}^2}$ \\ 
{\bf 3.} & $ x_{+}^2\in [0,\,1]$ and $ x_{-}^2<0$ & $[0,\,x_{+}]$ & $\sqrt{x_{+}^2+\left|x_{-}^2\right|}$ & 1 & $\frac{x_{+}^2}{x_{+}^2+\left|x_{-}^2\right|}$ \\ 
{\bf 4.}  & $x_{\pm}^2\in [0,\,1]$  & $[x_{-},\,x_{+}]$ & $x_{+}$ & 1 & $1-\frac{x_{-}^2}{x_{+}^2}$
\\ 
\end{tabular}
\end{ruledtabular}
\end{table} 
\noindent Case ({\bf 1}) corresponds to a circular string and the others to folded
strings, all along the $\xi$ coordinate.  We also see that every case has divergent energy
on the $x_{-}=0$, where $\sin\varphi=k=1$. In this limit we calculate the integrals necessary for the
deficit angles and for the momenta, 
\begin{eqnarray}
&&\intop_0^{x_+}\frac{dx}{(1-x)\sqrt{x_+^2-x^2}}=\frac{\pi-2\xi_+}{\sin
  2\xi_+},\qquad\intop_0^{x_+}\frac{dx}{(1+x)\sqrt{x_+^2-x^2}}=\frac{2\xi_+}{\sin2\xi_+},\nonumber\\
&&\intop_0^{x_+}\frac{xdx}{(1-x^2)\sqrt{x_+^2-x^2}}=\frac{\frac{\pi}{2}-2\xi_+}{\sin2\xi_+},\qquad\intop_0^{x_+}\frac{xdx}{\sqrt{x_+^2-x^2}}=\cos2\xi_+
\qquad\mbox{and}\nonumber \\
&&\intop_0^{x_+}\frac{dx}{\sqrt{x_+^2-x^2}}=\frac{\pi}{2},
\end{eqnarray}
\noindent where we used the definition $x=\cos2\xi$. $x_{-}=0$ also implies that
\begin{eqnarray}
x_{+}^2&=&2+\frac{\nu^2}{\omega^2}-\left(1+\frac{\beta^2}{\alpha^2}\right)\frac{\kappa^2}{\omega^2}\label{xp}\\
A_{\psi}&=&\frac{1}{4\left(\omega^2+\nu^2\right)\left(\alpha^2-\beta^2\right)}\left(-\beta\omega\kappa^2\pm\nu\sqrt{\left(\kappa^2-\nu^2-\omega^2\right)\left[\alpha^2\left(\omega^2+\nu^2\right)-\kappa^2\beta^2\right]}\,\right)\label{Apsi}.
\end{eqnarray}
%%%%%%%%%%%%
%
% Apsi, xp
%
$A_{\psi}$ must be real, and so we obtain the following relations among
the other parameters
\begin{equation}
\frac{\kappa^2}{\omega^2+\nu^2}\in\left[\,1,\,\frac{\alpha^2}{\beta^2}\right]\qquad
\mbox{if}\qquad \alpha^2>\beta^2,\qquad \mbox{and}\qquad\frac{\kappa^2}{\omega^2+\nu^2}\in\left[\,\,\frac{\alpha^2}{\beta^2},\,1\right]\qquad
\mbox{if} \qquad\beta^2>\alpha^2.
\end{equation}
If $\beta^2>\alpha^2$, then the velocity of the pulse is greather
than the velocity of the light, and so the the string has a tachyonic
character. Thus, we have characterized all the possible motions that a string
performs using (\ref{ansatz:1}) as ansatz. Now we consider the
particular cases where the energy, the momenta, and the deficit angle
are divergent.
\section{Particular Cases \label{s2}}
\subsection{Divergent Momenta and Finite Deficit Angles\label{MG}}
Looking at  (\ref{delta_phi}) and (\ref{delta_psi}) we can infer that by choosing
\begin{equation}
A_\psi=-\frac{1}{4}\frac{\omega\beta}{\alpha^2-\beta^2}\qquad
A=-\frac{1}{2}\frac{\nu\beta}{\alpha^2-\beta^2}
\end{equation}
\noindent we get finite deficit angles. Accordingly, the momenta are
\begin{equation}
 J_i=\frac{1}{2}\frac{\nu}{\kappa}E\mp \frac{T\pi}{4}\frac{\nu}{\omega}\qquad\mbox{and}\qquad
 J_\psi=\frac{\omega}{\kappa}E-T\cos\xi_+
\end{equation}
\noindent and the deficit angles are,
\begin{equation}
\Delta\phi_1=2\frac{\beta}{\alpha}\frac{\nu}{\omega}\frac{2\xi_+}{\sin2\xi_+},\qquad
\Delta\phi_2=2\frac{\beta}{\alpha}\frac{\nu}{\omega}\frac{2\xi_+ -\pi}{\sin2\xi_+},\qquad\mbox{and}\qquad
\Delta\psi=\frac{\beta}{\alpha}\frac{2\xi_+ -\frac{\pi}{2}}{\sin2\xi_+}.
\end{equation}
Looking at the difference between the Virasoro constraints (\ref{v1-v2}) we also
obtain a relation among the constant parameters
\begin{equation}
\kappa^2=\omega^2+\nu^2,
\end{equation}
\noindent which is consistent with (\ref{Apsi}) and implies that
\begin{equation}
\cos^22\xi_+=x_+^2=1-\frac{\beta^2}{\alpha^2}\left(1+\frac{\nu^2}{\omega^2}\right),
\end{equation}
\noindent The choice $\nu=0$ leads to
\begin{equation}
J_i=\Delta\phi_i=0, \qquad \sin2\xi_+=\frac{\beta}{\alpha},\qquad\Delta\psi=2\xi_+-\frac{\pi}{2},
\end{equation}
\noindent and we get the dispersion relation of a giant magnon in the
coordinate $\psi$ 
\begin{equation}
E-J_\psi=-T\sin\Delta\psi.\label{mg_A}
\end{equation}
\noindent This is expected, because the choice $\nu=0$ restricts the movement of
the string to an $\mathbb{R}\times S^2$ sector. On the other hand,
$\nu\neq 0$ keeps the string with three angular momenta. Defining $J=J_1+J_2$ and $j=J_1-J_2$, we write the dispersion relation
\begin{equation}
\sqrt{E^2-J^2}-J_{\psi}=-T\sin\left(\sqrt{1+\frac{4j^2}{T^2\pi^2}}\Delta\psi\right),\label{DR}
\end{equation}
%%%%%%%%%%
%
% DR
%
%%
\noindent where 
\begin{equation}
\frac{\nu}{\omega}=\frac{\Delta\phi_1+\Delta\phi_2}{4\Delta\psi}=-\frac{2j}{T\pi}\label{D1_D2}
\end{equation}
\noindent was used. The question which arises is whether the dispersion relation
(\ref{DR}) describes a genuine
giant magnon with three angular momenta, and we hypotesize that it does. Dispersion relations like (\ref{DR}) have already been used to
describe a giant magnon with two angular momenta in
\cite{Kluson:2007qu,Ryang:2008rc}, and the novelty in the above
relation is the square root multiplying $\Delta\psi$ in the argument of
the sine. However, imposing the additional constraint
\begin{equation}
\sqrt{1+\frac{4j^2}{T^2\pi^2}}=1+\frac{n\pi}{\Delta\psi},\qquad
\mbox{with}\qquad n\in\mathbb{N}
\end{equation}
\noindent results in the expected relation, and of course $n=0$ returns
(\ref{mg_A}). This interesting constraint between $j$ and $\Delta\psi$
seems to quantize its relation in order to produce a giant
magnon. Nevertheless, if this quantizing 
constraint is not imposed, one hypothesis can immediately
be raised: as the sine argument is always greater than $\Delta\psi$, it is
possible for the deficit angle to be a lower bound, and so deficit
angles greater than $\Delta\psi$ could represent a giant magnon. 
Of course, these possibilities need a careful study on the gauge side
of the duality, and we hope this result could be used to clarify  this
point in the future.
\subsection{Finite Momenta and Divergent Deficit Angles}
To this case the choice 
\begin{equation}
A=-\frac{1}{2\beta}\frac{\nu\alpha^2}{\alpha^2-\beta^2}\qquad
A_\psi=-\frac{1}{4\beta}\frac{\omega\alpha^2}{\alpha^2-\beta^2},
\end{equation}
\noindent in  (\ref{Apsi}) and (\ref{xp}) implies that
\begin{equation}
\alpha^2\left(\omega^2+\nu^2\right)-\kappa^2\beta^2=0\qquad\mbox{and}\qquad x_{+}^2=1-\frac{\alpha^2}{\beta^2}\left(1+\frac{\nu^2}{\omega^2}\right),\label{tachyon}
\end{equation}
\noindent  and the momenta are
\begin{equation}
J_i=\mp\frac{T}{4}\frac{\nu}{\omega}\qquad\mbox{and}\qquad J_\psi=-Tx_+.
\end{equation}
From these relations, we can write
\begin{equation}
\frac{\beta^2}{\alpha^2}=\frac{T^2+4j^2}{T^2-J_\psi^2},\label{nao}
\end{equation}
\noindent which shows the dependence between the velocity of the pulse
on the string and the angular momenta on the coordinates. 
From (\ref{tachyon}) we see that $\beta^2>\alpha^2$, and so
this string is tachyonic. A tachyon like this also appears in the case of a
spiky string with two angular momenta in this background
\cite{Ryang:2008rc}. Now, we write the deficit angles 
\begin{eqnarray}
&&\Delta\phi_1=2\frac{\alpha}{\beta}\frac{\nu}{\omega}\frac{2\xi_+}{\sin2\xi_+}-\frac{\nu}{\sqrt{\omega^2+\nu^2}}\frac{E}{T}\\
&&\Delta\phi_2=2\frac{\alpha}{\beta}\frac{\nu}{\omega}\frac{2\xi_+-\pi}{\sin2\xi_+}-\frac{\nu}{\sqrt{\omega^2+\nu^2}}\frac{E}{T}\\
&&\Delta\psi=\frac{\alpha}{\beta}\frac{2\xi_+-\frac{\pi}{2}}{\sin2\xi_+}-\frac{\omega}{\sqrt{\omega^2+\nu^2}}\frac{E}{T}.
\end{eqnarray}
\noindent $\nu=0$ generates the relation expected for a spiky string
in one dimension 
\begin{equation}
E+T\Delta\psi=T\left(2\xi_+-\frac{\pi}{2}\right).\label{SS1}
\end{equation}
\noindent which allows us to write the dispersion relation
\begin{equation}
E+\sqrt{T^2+4j^2}\left[\Delta\psi+\frac{T}{8j}\left(\Delta\phi_2-\Delta\phi_1\right)\right]=T\,2\xi_+, \label{SS}
\end{equation}
\noindent which gives the spiky string with one angular momentum (\ref{SS1}) in
$j\to 0$. The case is similar to the case of the giant magnon of the
previous item: we have a dispersion relation that is not exactly the
expected one but with a well-known one dimensional limit. As in the
former case, we use this evidence to claim that (\ref{SS}) represents
a spiky string, and we envisage that further studies on the gauge
sector of the correspondence will confirm it. 
\subsection{Finite $J_\psi$ and Divergent $J_i$}
We can also study cases that do not have both the divergent momenta or
both the finite momenta. If we have $J_\psi$ finite and $J_i$ divergent, the choice
\begin{equation}
A_\psi=-\frac{1}{4\beta}\frac{\omega\alpha^2}{\alpha^2-\beta^2},\qquad
A=-\frac{1}{2}\frac{\nu\beta}{\alpha^2-\beta^2},
\end{equation}
\noindent and (\ref{v1-v2}), (\ref{Apsi}), and (\ref{xp}) give
\begin{equation}
\kappa^2=\frac{\alpha^2}{\beta^2}\omega^2+\nu^2\qquad
\mbox{and}\qquad x_+^2=1-\frac{\alpha^2}{\beta^2}-\frac{\beta^2}{\alpha^2}\frac{\nu^2}{\omega^2},
\end{equation}
\noindent which implies in $\beta^2>\alpha^2$, and $\omega^2>\nu^2$,
which means that the configuration is tachyonic, as was found in the
preceding case. The momenta in this case are
\begin{equation}
J_i=\frac{1}{2\sqrt{1+\frac{\alpha^2}{\beta^2}\frac{\omega^2}{\nu^2}}}E\mp\frac{T\pi}{4}\frac{\nu}{\omega}\qquad\mbox{and}\qquad J_\psi=-Tx_+.
\end{equation}
\noindent The deficit angles read
\begin{eqnarray}
&&\Delta\phi_1=\frac{\beta}{\alpha}\frac{\nu}{\omega}\frac{4\xi_+}{\sin2\xi_+},\qquad\Delta\phi_2=\Delta\phi_1-\frac{\beta}{\alpha}\frac{\nu}{\omega}\frac{2\pi}{\sin2\xi_+},\qquad\mbox{and}
\nonumber \\
 && \Delta\psi=\frac{1}{2}\frac{\omega}{\nu}\Delta\phi_2-\frac{E}{T\sqrt{1+\frac{\beta^2}{\alpha^2}\frac{\nu^2}{\omega^2}}}.
\end{eqnarray}
\noindent A giant magnon is not possible in this case, because it
would need a $\omega\to 0$. On the other hand, looking at the deficit angles, we write
\begin{equation}
\sqrt{E^2-J^2}+T\Delta\psi=-\frac{T^2\pi}{4}\frac{\Delta\phi_1+\Delta\phi_2}{j}\label{SS_C},
\end{equation}
%%%%%%%%
%
% SS_C
%
\noindent which is not exactly an expected relation for a spiky string, but
it has the same structure as the spiky string with two angular momenta found by Ryang
\cite{Ryang:2008rc}. The right hand side of (\ref{SS_C}) is given in
terms of deficit angles in the $\phi_i$ directions, which is
unexpected, as the string peaks along the $\xi$ direction. Yet, choice $\nu=0$ leads to
\begin{equation}
\frac{\kappa^2}{\omega^2}=\frac{\alpha^2}{\beta^2}=\sin^22\xi_+
\end{equation}
\noindent and consequently we can write the relation
\begin{equation}
E+T\Delta\psi=T\left(4\xi_+-\pi\right),
\end{equation}
\noindent which shows that there is a one-dimensional spiky string
limit in the configuration. Indeed (\ref{SS_C}) is telling us that
there is a constraint among the deficit angles in the $\phi_i$
directions and the peak in the $\xi$ direction. This interpretation 
conforms with the former cases, where new constraints were found,
and a well-behaved one-dimensional limit can be reached. 
\subsection{Finite $J_i$ and Divergent $J_\psi$}
In this case the integration constants of the problem are
\begin{eqnarray}
 A=-\frac{1}{2\beta}\frac{\nu\alpha^2}{\alpha^2-\beta^2},\qquad
A_\psi=-\frac{1}{4}\frac{\omega\beta}{\alpha^2-\beta^2}
\end{eqnarray}
\noindent which, using  (\ref{Apsi}) and (\ref{xp}), imply that
\begin{equation}
\kappa^2=\omega^2+\frac{\alpha^2}{\beta^2}\nu^2\qquad\mbox{and}
\qquad x_+^2=1-\frac{\beta^2}{\alpha^2}-\frac{\alpha^2}{\beta^2}\frac{\nu^2}{\omega^2}
\end{equation}
\noindent which means that $\alpha^2>\beta^2$, and
$\omega^2>\nu^2$. Accordingly, we calculate the momenta
\begin{equation}
J_i=\mp\frac{\nu}{\omega}\frac{T\pi}{4},\qquad\mbox{and}\qquad J_\psi=\frac{1}{\sqrt{1+\frac{\alpha^2}{\beta^2}\frac{\nu^2}{\omega^2}}}E-Tx_+,
\end{equation}
\noindent and the deficit angles
\begin{eqnarray}
&&\Delta\phi_1=\frac{\alpha}{\beta}\frac{\nu}{\omega}\frac{4\xi_+}{\sin2\xi_+}-\frac{1}{2T}\frac{E}{\sqrt{1+\frac{\beta^2}{\alpha^2}\frac{\omega^2}{\nu^2}}}\qquad
\Delta\phi_2=\Delta\phi_1-\frac{\alpha}{\beta}\frac{\nu}{\omega}\frac{2\pi}{\sin2\xi_+} \\
&&\Delta\psi=\frac{\beta}{\alpha}\frac{\frac{\pi}{2}-2\xi_+}{\sin2\xi_+}\qquad\mbox{and}\qquad
\Delta\phi_1-\Delta\phi_2=\frac{\alpha^2}{\beta^2}\frac{j}{T}\frac{\Delta\psi}{2\xi_+-\frac{\pi}{2}}.
\end{eqnarray}
This is probably the most unusual of the four cases studied here. Now
we have a $\nu=0$ limit where a giant magnon is found. On the other
hand, if $\nu\neq 0$ there is a spiky string-like dispersion relation
involving the $\phi_i$ angles. Our interpretation is that we have the
giant magnon and the spiky string coexisting and with a coupled
dispersion relationd, namely
\begin{eqnarray}
&&\frac{E}{T}-\sqrt{1+\frac{\beta^2}{\alpha^2}\frac{T^2\pi^2}{4j^2}}\,\frac{\Delta\phi_1+\Delta\phi_2}{2}=\frac{1}{2}\frac{\alpha}{\beta}\sqrt{1+\frac{\alpha^2}{\beta^2}\frac{4j^2}{T^2\pi^2}}\Delta\psi\\
&& \nonumber\\
&&\frac{E}{\sqrt{1+\frac{\alpha^2}{\beta^2}\frac{4j^2}{T^2\pi^2}}}-J_\psi=T\cos2\xi_+
\end{eqnarray}
This is a new kind of structure and, of course, it requires further
studies for a better understanding. 

\section{Conclusion\label{s3}}
In this article divergent energy classical strings with three angular
momenta were studied. There is evidence that these strings
actually correspond to the until-now-undiscovered spiky string
and giant magnon configurations, although this evidence is  not conclusive. On the other hand, these structures
do not obey the usual prescriptions which are known for
an $\mathbb{R}\times S^2$ subspace or even to strings moving in higher
dimensional  spherically symmetrical spaces. As the
$\mathbb{R}\times\mathbb{CP}^3$ is something different from $\mathbb{R}\times S^n$ 
we expect different dispersion relations, at least while there are not, to our
knowledge, gauge studies that confirm or deny this hypothesis.

{\bf Acknowledgements}

Sergio Giardino's research activities have the support of CNPq and Hector Carri\'{o}n is supported
by FAPESP and PROSUL. Hector
Carri\'{o}n thanks the Centro Brasileiro de Pesquisas F\'{i}sicas
for their hospitality during the short period he was a
research fellow there.  The authors are also grateful to Leandro
Ibiapina Belvil\'{a}qua and Victor O. Rivelles for their input.

%%%%%%%%%%%%%%%%%%%%%%%%
%
%
%  BIBLIOGRAPHY
%
%
\bibliographystyle{unsrt} 
\bibliography{bib_B0}
%%%%%%%%%%%%%%%%%%
%
%
%    END
%
%
\end{document}